\documentclass[aps,prl,twocolumn,amsmath,amssymb,showpacs]{revtex4-1}
\usepackage[ansinew]{inputenc}
\usepackage{graphicx}
\usepackage{textcomp}
\usepackage{bbm}
\usepackage{hyperref}

\newcommand{\bra}[1]{\langle#1|}
\newcommand{\ket}[1]{|#1\rangle}
\newcommand{\Fidelity}{\mathcal{F}}
\newcommand{\FidelityEnt}{\Fidelity_\mathrm{ent}}
\newcommand{\FidelityA}{\Fidelity_\mathrm{A}}
\newcommand{\FidelityOrth}{\Fidelity_\mathrm{\perp}}
\newcommand{\FidelityBell}{\Fidelity_\mathrm{\|}}
\newcommand{\rhoEnt}{\rho_\textnormal{ent}}
\newcommand{\pEnt}{p_\textnormal{ent}}
\newcommand{\pA}{p_\mathrm{A}}
\newcommand{\spinup}{\ket{\!\uparrow}}
\newcommand{\spindown}{\ket{\!\downarrow}}
\newcommand{\spinupx}{\ket{\!\uparrow_\mathrm{x}}}
\newcommand{\spindownx}{\ket{\!\downarrow_\mathrm{x}}}
\newcommand{\spinupy}{\ket{\!\uparrow_\mathrm{y}}}
\newcommand{\spindowny}{\ket{\!\downarrow_\mathrm{y}}}
\newcommand{\photonright}{\ket{\!\circlearrowright}}
\newcommand{\photonleft}{\ket{\!\circlearrowleft}}
\newcommand{\pw}{p_\mathrm{w}}

\begin{document}
\title{Efficient Teleportation between Remote Single-Atom Quantum Memories}

\author{Christian N\"{o}lleke}
\author{Andreas Neuzner}
\author{Andreas Reiserer}
\author{Carolin Hahn}
\author{Gerhard Rempe}
\author{Stephan Ritter}
\email{stephan.ritter@mpq.mpg.de}

\affiliation{Max-Planck-Institut f\"{u}r Quantenoptik, Hans-Kopfermann-Strasse 1, 85748 Garching, Germany}

\begin{abstract}
We demonstrate teleportation of quantum bits between two single atoms in distant laboratories. Using a time-resolved photonic Bell-state measurement, we achieve a teleportation fidelity of $(88.0\pm1.5)\,\%$, largely determined by our entanglement fidelity. The low photon collection efficiency in free space is overcome by trapping each atom in an optical cavity. The resulting success probability of $0.1\,\%$ is almost 5 orders of magnitude larger than in previous experiments with remote material qubits. It is mainly limited by photon propagation and detection losses and can be enhanced with a cavity-based deterministic Bell-state measurement.
\end{abstract}

\maketitle

The faithful transfer of quantum information between distant memories that form the nodes of a quantum network is a major goal in applied quantum science \cite{Kimble2008}. One way to achieve this is via direct transfer, e.g., by the coherent exchange of a single photon \cite{Ritter2012}. Over large distances, however, the inevitable losses in any quantum channel render this scenario unrealistic, as its efficiency decreases exponentially with the distance between the network nodes. For any classical information, the solution is simple: It can be amplified at intermediate nodes of the network. It can also be copied before transmission, allowing for a new transmission attempt should the previous one have failed. For a quantum state, the no-cloning theorem states that this is impossible. Therefore, quantum repeater schemes have been proposed to establish long-distance entanglement using photons and memories \cite{Briegel1998, Duan2001}. This entanglement can then be used as a resource for the transfer of quantum information via teleportation \cite{Bennett1993}.

The underlying principle of teleportation was first realized with photonic qubits \cite{Bouwmeester1997,Boschi1998,Furusawa1998} and since then has been exploited in many experiments \cite{Yin2012,Ma2012}. Teleportation between matter qubits was first achieved with trapped ions \cite{Riebe2004,Barrett2004}, albeit over a distance limited to a few micrometers owing to the short-range Coulomb interaction. Teleportation between distant material qubits, however, requires photons distributing entanglement, as was demonstrated with two single ions separated by about 1\,m \cite{Olmschenk2009}. The low photon-collection efficiency in free space, however, prevents scaling of that approach to larger networks. We eliminate this obstacle by trapping two remote single atoms each in an optical cavity. This allows for an in principle deterministic creation of atom-photon entanglement and atom-to-photon state mapping using a vacuum-stimulated Raman adiabatic passage (vSTIRAP) technique \cite{Kuhn2002, Wilk2007, Weber2009}. To teleport the stationary qubit at the sender atom, encoded in two Zeeman states of the atomic ground-state manifold, we map it onto a photonic qubit and perform a Bell-state measurement (BSM) between this photon and that of an entangled atom-photon state originating from the receiver atom \cite{Weinfurter1994,Braunstein1995}. Compared to realizations with atoms in free space, the use of cavities boosts the overall efficiency by almost 5 orders of magnitude \cite{Olmschenk2009}.

\begin{figure}
\centering
\includegraphics[width=\columnwidth]{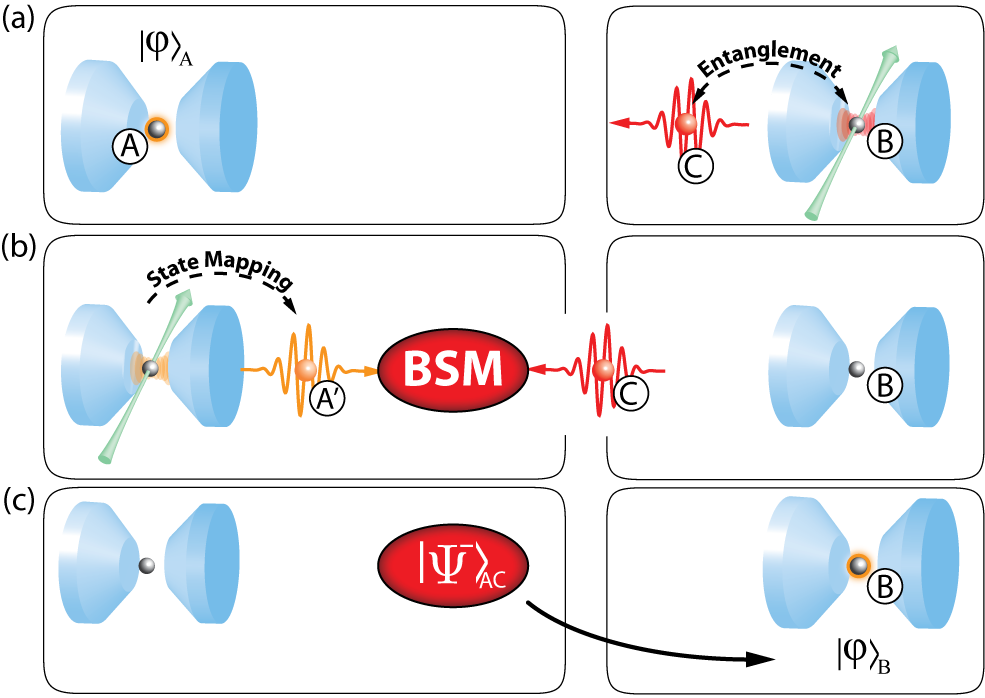}
\caption{
Experimental setup for teleportation between remote single-atom quantum memories. Single atoms (grey spheres A and B) are trapped in optical cavities (blue cones) in independent laboratories at a distance of 21m. (a) Entanglement is generated between atom B and an ancilla photon C. (b) The atomic qubit at node A is mapped onto a photonic qubit A$^\prime$ and a Bell-state measurement between the two photons is performed. (c) Detection of a $\ket{\Psi^-}$ event heralds the successful teleportation of the atomic qubit from node A to node B.
\label{fig:setup}}
\end{figure}

In our experiment, single $^{87}$Rb atoms trapped in high-finesse optical cavities act as quantum memories at both node A and node B (Fig.\,\ref{fig:setup}). The independent systems have a distance of 21\,m and operate in the intermediate coupling regime of cavity QED. Quasipermanent trapping is achieved by using far-off-resonance dipole traps that shift the relevant atomic transition frequency by 150\,MHz (node A) and 115\,MHz (node B), respectively. We identify the states $\spindown$ and $\spinup$ with the atomic $\ket{F,m_F}$ ground states $\spindown_\mathrm{A} = \ket{2,-1}_\mathrm{A}$, $\spinup_\mathrm{A} = \ket{2,+1}_\mathrm{A}$ and $\spindown_\mathrm{B} = \ket{1,-1}_\mathrm{B}$, $\spinup_\mathrm{B} = \ket{1,+1}_\mathrm{B}$, where the index denotes the memory atom at node A and B, respectively.

To demonstrate teleportation between the two memories, we initialize node A in the state $\ket{\varphi}_\mathrm{A}=\alpha\spindown+\beta\spinup$ by setting the polarization of a weak coherent laser pulse (about five photons on average) and mapping it onto the atomic spin with the procedure used in \cite{Specht2011}. At node B, entanglement is created locally between the spin state of the atom and the polarization of a photon C using a vSTIRAP \cite{Wilk2007, Weber2009} [Fig.\,\ref{fig:setup}(a)]. The maximally entangled atom-photon state reads
\begin{equation}
\label{eq:Psim_exp}
\ket{\Psi^-}_\mathrm{BC}=\frac{1}{\sqrt{2}}\left(\spindown_\mathrm{B}\photonright_\mathrm{C} - \spinup_\mathrm{B}\photonleft_\mathrm{C}\right).
\end{equation}
Here $\photonright$ and $\photonleft$ denote right- and left-circularly polarized photon states, respectively. The subscripts label the individual particles. Photon C is sent to memory A via an optical fiber. The three-particle state of the qubits A, B and C can be written as
\begin{align}
\label{eq:Teleportation}
\ket{\varphi}_\mathrm{A}\ket{\Psi^-}_\mathrm{BC} =  &\frac{1}{2}\left(\ket{\Phi^+}_\mathrm{AC}\hat{\sigma}_\mathrm{x}\hat{\sigma}_\mathrm{z}\ket{\varphi}_\mathrm{B}-\ket{\Phi^-}_\mathrm{AC}\hat{\sigma}_\mathrm{z}\ket{\varphi}_\mathrm{B} \right. \nonumber \\
&\left.+\ket{\Psi^+}_\mathrm{AC}\hat{\sigma}_\mathrm{x}\ket{\varphi}_\mathrm{B}-\ket{\Psi^-}_\mathrm{AC}\ket{\varphi}_\mathrm{B}\right),
\end{align}
$\ket{\Phi^\pm}$ and $\ket{\Psi^\pm}$ denote the four normalized, maximally entangled Bell states in the basis of our BSM setup, and $\hat{\sigma}_i$ denote the Pauli operators ($i= \mathrm{x,y,z}$). A BSM of the qubits A and C projects qubit B onto the initial state $\ket{\varphi}$ of qubit A up to a unitary transformation that depends on the measurement outcome. We perform the BSM optically, by employing single-photon detectors and linear-optics quantum interference between the two photons \cite{Weinfurter1994,Braunstein1995}. For this purpose we first map the atomic state of memory A onto the polarization of a photon A$^\prime$ [Fig.\,\ref{fig:setup}(b)]. Then the two photons A$^\prime$ and C are superimposed on a nonpolarizing beam splitter (NPBS). We adjust the control laser frequency and intensity of the vSTIRAP aiming for identical frequency and wave packet envelope of the photons \cite{sm}. They will always be detected in the same output port if their common wave function is symmetric with respect to particle exchange, and in different ports if it is antisymmetric. This allows for unambiguous detection of the $\ket{\Psi^-}$ Bell state which directly heralds the successful state transfer from memory A to memory B [Fig.\,\ref{fig:setup}(c)]. In combination with polarization-sensitive detection the $\ket{\Psi^+}$  Bell state can be identified as well. The simplicity of the optical BSM comes at the price of an inherently probabilistic, but nonetheless heralded, process, with a maximum efficiency of 0.5, as only two out of the four photonic Bell states can be identified unambiguously \cite{Calsamiglia2001}.

The photon-production efficiency into the single spatial mode defined by the cavity is 39\,\% at node A and 25\,\% at node B. These photons are detected in the BSM setup with an efficiency of 31\,\% and 12\,\%, respectively. These numbers include all propagation losses and the quantum efficiency of the detectors \cite{sm}, such that the probability for a two-photon correlation is 0.36\,\%. Our teleportation is conditioned on the detection of a $\ket{\Psi^-}$ Bell state. As all four Bell states are equally probable ($1/4$ each), the success probability of teleportation is 0.1\,\%. In contrast to previous demonstrations, the efficiency is therefore not predominantly limited by the single-photon generation and collection efficiency but by the requirement to transmit and detect two photons simultaneously, which is inherent in the optical BSM.

\begin{table}
\centering
\begin{tabular} {l|c}
\textbf{Input state} & \textbf{Fidelity (\%)}\\\hline\hline
$\spindown$ & $74.5\pm2.6$\\\hline
$\spinup$ & $72.3\pm2.8$\\\hline
$\spindowny = \frac{1}{\sqrt{2}}\left(\spindown + i \spinup\right)$ & $73.0\pm3.0$\\\hline
$\spinupy = \frac{1}{\sqrt{2}}\left(\spindown - i \spinup\right)$ & $75.0\pm3.0$\\\hline
$\spindownx = \frac{1}{\sqrt{2}}\left(\spindown + \spinup\right)$ & $88.6\pm2.3$\\\hline
$\spinupx = \frac{1}{\sqrt{2}}\left(\spindown - \spinup\right)$ & $90.2\pm2.5$\\\hline\hline
Average & $\mathbf{78.9\pm1.1}$\\
\end{tabular}
\caption{
Teleportation fidelity conditioned on a $\ket{\Psi^-}$ detection event. The table shows the individual teleportation fidelities, defined as the overlap between the state ideally prepared at node A (input state) and the teleported state at node B, for six mutually unbiased input states of node A. The quoted errors are the statistical standard error.
\label{table:fidelity}}
\end{table}

\begin{figure*}
\centering
\includegraphics[width=\textwidth]{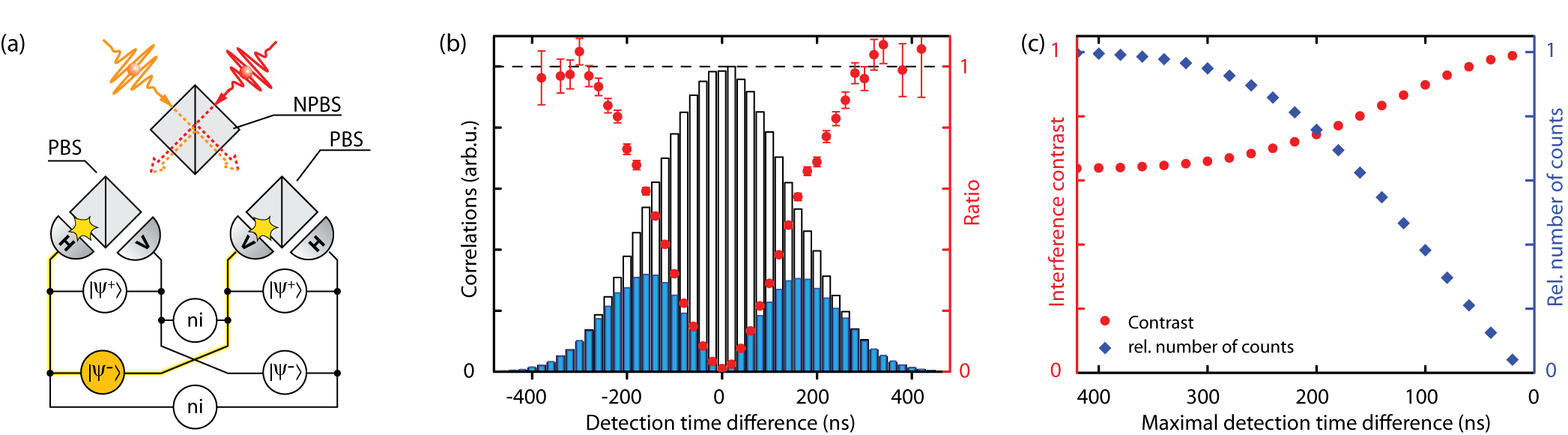}
\caption{
Bell-state measurement using time-resolved two-photon interference. (a) The photons (red and orange) overlap at an NPBS. Two single-photon detectors (gray semicircles) and a polarizing beam splitter (PBS) in each output port of the NPBS allow us to discriminate between horizontally (H) and vertically (V) polarized photons. Correlations between different detectors either herald a $\ket{\Psi^-}$ or a $\ket{\Psi^+}$ Bell state or indicate that the photons did not interfere (ni). (b) Result of a time-resolved coincidence measurement for photons of identical ($N_\mathrm{ni}$, blue filled bars) and orthogonal ($N_{\ket{\Psi^-}}$, black open bars) polarization in opposite output ports of the NPBS. The number of coincidences between photons with the same polarization is strongly suppressed. The red data points show the ratio $N_\mathrm{ni}/N_{\ket{\Psi^-}}$ between no-interference and $\ket{\Psi^-}$ events. The error bars indicate the statistical standard error. (c) Interference contrast $C$ (red points) and relative number of evaluated counts (blue diamonds) as a function of the maximal detection time difference for a two-photon correlation in different output ports of the NPBS. By postselecting on short detection time differences, the interference contrast can be greatly enhanced. All statistical error bars are smaller than the respective symbols.
\label{fig:BSM}}
\end{figure*}

In order to characterize the fidelity of the teleportation, we perform quantum state tomography on atom B conditioned on a $\ket{\Psi^-}$ detection \cite{Nielsen2000}. For this purpose, the state of atom B is mapped onto the polarization of another photon B$^\prime$, whose polarization is measured. The fidelity is defined as the overlap between the state $\ket{\varphi}$ ideally prepared at node A and the density matrix $\rho_\mathrm{B^\prime}$ of photon B$^\prime$ following a successful $\ket{\Psi^-}$ detection: $\Fidelity = \bra{\varphi} \rho_\mathrm{B^\prime} \ket{\varphi}$. The measured fidelity thus includes imperfections during state preparation and state readout and is therefore a lower bound to the fidelity of the teleportation itself \cite{sm}.

Table \ref{table:fidelity} shows the fidelity for six input states initially prepared at node A, forming three mutually unbiased bases. The fidelity for input states which are mapped onto eigenpolarizations of the detection basis of the BSM setup ($\spindownx$ and $\spinupx$) is considerably larger than for the eigenstates of the other bases, because it does not depend on the quality of the two-photon interference \cite{sm}. For the average fidelity, defined as the mean of the six individual values, we find $\Fidelity = (78.9\pm 1.1)\,\%$, more than 10 standard deviations above the classical limit of $2/3$. In case a $\ket{\Psi^+}$ Bell state is detected in the experiment, node B is projected to a state that is rotated with respect to the input state [Eq.\,(\ref{eq:Teleportation})]. With respect to this rotated state $\hat{\sigma}_\mathrm{x}\ket{\varphi}$, we find $\Fidelity_{\Psi^+}=(82.4\pm 1.1)\,\%$. All quoted uncertainties reflect statistical standard errors.

The teleportation fidelity directly depends on the contrast $C$ of the two-photon interference used to implement the Bell-state measurement, the entanglement fidelity $\FidelityEnt$, and the fidelity of state preparation and mapping $\FidelityA$ at atom A. Using a simple model we find \cite{sm}
\begin{equation}
\label{eq:ExpectedFidelityT}
\Fidelity = \frac{1}{2} + \frac{8}{9} \left( C + \frac{1}{2} \right) \left( \FidelityEnt - \frac{1}{4} \right) \left( \FidelityA - \frac{1}{2} \right).
\end{equation}
The entanglement fidelity $\FidelityEnt$ is defined as the overlap between the readout state of atom B and photon C with the ideal $\ket{\Psi^-}$ Bell state.

The achieved interference contrast $C$ determines the quality of the optical BSM. Assuming perfect interference, only photons in the $\ket{\Psi^-}$ state will lead to correlations in different output ports of the NPBS [Fig.\,\ref{fig:BSM}(a)]. In this state, the photons have orthogonal polarization. Therefore, the number of coincidences between detectors in different output ports of the NPBS and with identical polarization $N_\mathrm{ni}$ should be zero. Its ratio to the number $N_{\ket{\Psi^-}}$ of $\ket{\Psi^-}$ events is thus a direct measure of the distinguishability of the generated photons.

For the interference contrast, defined as $C=1-N_\mathrm{ni}/N_{\ket{\Psi^-}}$, we measure $C=64\,\%$, clearly demonstrating quantum interference between the two photons emitted from the independent memories. We attribute the nonperfect contrast to fluctuations within the photonic wave packets, most likely caused by the uncorrelated motions of the two trapped atoms that lead to fluctuating ac Stark shifts and fluctuating atom-cavity coupling strength. In this context, using a vSTIRAP for photon production has two advantages. First, the frequency of the emitted photons can be tuned. This allows for the creation of frequency-matched photons even though the relevant atomic transition frequencies at the two nodes differ by about six atomic linewidths due to different ac Stark shifts induced by the used dipole traps \cite{sm}. Second, it allows us to generate photons with a temporal length ($\simeq 250$\,ns) largely exceeding the temporal resolution of our detection system ($\leq5$\,ns). This enables us to herald teleportation events with increased fidelity. For this purpose we investigate the contrast of two-photon interference as a function of the detection time difference \cite{Legero2004}. The correlations between the two NPBS output ports, $N_{\ket{\Psi^-}}$ and $N_\mathrm{ni}$ are plotted in Fig.\,\ref{fig:BSM}(b). For short time differences $N_\mathrm{ni}$ nearly vanishes, providing near-perfect discrimination of $\ket{\Psi^-}$. This proves that the BSM setup is well aligned and that the single-photon sources exhibit excellent antibunching. Reducing the coincidence time window allows us to increase the interference contrast to almost unity, while the success probability of the whole protocol naturally decreases [Fig.\,\ref{fig:BSM}(c)]. For time differences shorter than 20\,ns, the contrast is 98.9\,\%.

\begin{figure}
\centering
\includegraphics[width=\columnwidth]{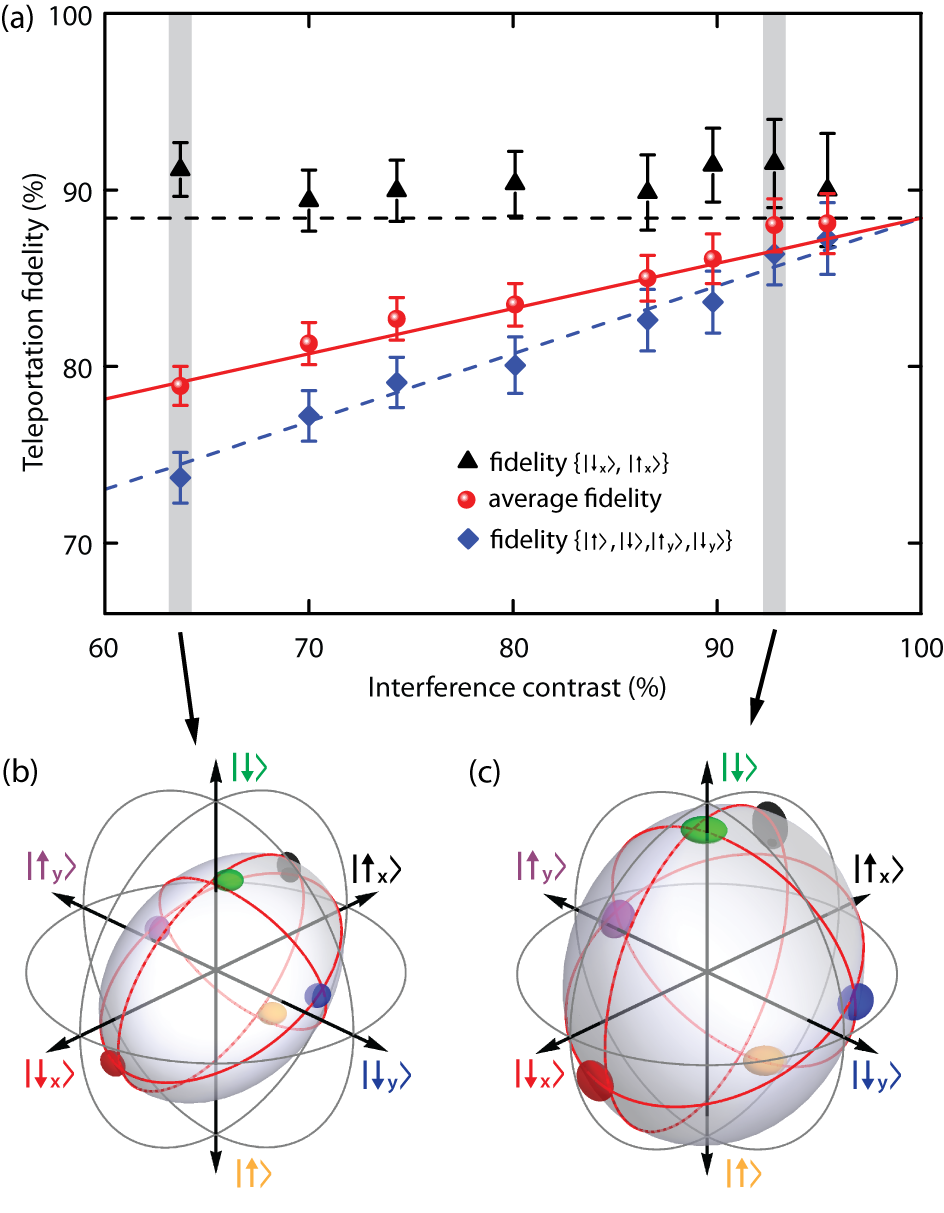}
\caption{
Teleportation fidelity as a function of the interference contrast. Increasing the contrast by using time-resolved BSM (Fig.\,\ref{fig:BSM}) increases the fidelity of the whole teleportation protocol. (a) The plot shows data for $\ket{\Psi^-}$ teleportation events. Black triangles are measured average fidelities for the two input states $\spindownx$ and $\spinupx$. The blue diamonds are averages over the other four input states given in Table \ref{table:fidelity}. The red points show the mean fidelity averaged over all six input states. The red line is the theoretical expectation [Eq.\,(\ref{eq:ExpectedFidelityT})] for the parameters $\FidelityEnt=89\,\%$ and $\FidelityA=95\,\%$ estimated from independent measurements and including imperfections in the readout \cite{sm}. The dashed lines indicate the theoretically expected behavior for the subset of data in the respective color \cite{sm}. The error bars indicate the statistical standard error. (b), (c) Reconstructed Bloch sphere \cite{Specht2011} of the teleported state, with no additional temporal filtering applied ($C=64\,\%$) (b) and for a contrast of $C=93\,\%$, i.e. only taking correlations with a maximal detection time difference of 80\,ns during the BSM (c). The size of the colored ellipsoids represents the statistical standard error.
\label{fig:Fidelity-Contrast}}
\end{figure}

We now apply the criterion of short detection time differences as an additional heralding condition for teleportation events. This allows us to dramatically increase the teleportation fidelity. Figure \ref{fig:Fidelity-Contrast}(a) shows the measured teleportation fidelity as a function of the interference contrast for different data subsets. Solid and dashed lines are theory plots [Eq.\,(\ref{eq:ExpectedFidelityT}) and Ref.\,\cite{sm}, respectively] according to our simple model. The average fidelity (red points) exhibits good agreement with the linear relation expected from Eq.\,(\ref{eq:ExpectedFidelityT}) (red line). As an example, reducing the coincidence time window to 80\,ns increases the contrast to $92.8\,\%$ and the teleportation fidelity to $(88.0\pm 1.5)\,\%$ for a detected $\ket{\Psi^-}$ correlation (Fig.\,\ref{fig:Fidelity-Contrast}). In the case of the two input states $\spindownx$ and $\spinupx$ there is no need to detect the symmetry of the state, as polarization correlations between photons that did not interfere are sufficient to herald the projection of atom B into the state initially prepared at atom A \cite{sm}. The measured fidelity is thus independent of the interference contrast $C$ [black triangles in Fig.\,\ref{fig:Fidelity-Contrast}(a)] and always higher than for all other states. This results in the ellipticity of the reconstructed Bloch sphere of the teleported state [Fig.\,\ref{fig:Fidelity-Contrast}(b)].

The demonstrated ability to teleport a quantum state between nonidentical memories opens up new perspectives for solid-state-based approaches to quantum networks, where identical network nodes are hard to realize \cite{Patel2010,Lettow2010,Bernien2012,Sipahigil2012}. Moreover, with the use of optical cavities we have put teleportation between material systems into a regime where the time needed for a successful teleportation event (about 0.1\,s at a repetition rate of 10\,kHz) is shorter than the coherence times observed in single atoms \cite{Langer2005}.

The optical BSM can also be applied to perform entanglement swapping \cite{moehring2007,Hofmann2012} as required for the experimental realization of a quantum repeater. Nevertheless, it comprises two efficiency limits. One is a fundamental upper bound of $1/2$, as only two of the four Bell states can be identified unambiguously \cite{Calsamiglia2001}. The other is the requirement to efficiently detect two single photons. Our cavity-based approach provides the possibility to overcome both bottlenecks in the future: The interaction of two atoms with one cavity mode \cite{Cirac1994,Lloyd2001} and atomic state detection \cite{Bochmann2010,Gehr2010} can be used for a BSM that has the potential to be deterministic, as it discriminates all four Bell states and does not require single-photon detection. This paves the way for more complex quantum networks with many nodes.

\textit{Note added.}---After submission of this paper we became aware of related work with atomic ensembles \cite{Bao2012}.

\begin{acknowledgments}
We thank David Moehring for contributions during the early stage of the experiments and Eden Figueroa and Manuel Uphoff for discussions. This work was supported by the Deutsche Forschungsgemeinschaft (Research Unit 635), by the European Union (Collaborative Project AQUTE) and by the Bundesministerium für Bildung und Forschung via IKT 2020 (QK\_QuOReP).
\end{acknowledgments}

\setcounter{figure}{0}
\renewcommand{\thefigure}{S\arabic{figure}}
\setcounter{equation}{0}
\renewcommand{\theequation}{S\arabic{equation}}

\section*{Methods}
\subsection*{Experimental setup}
The two independent quantum nodes are designed to operate with similar physical parameters. In each apparatus, a single $^{87}$Rb atom is quasi-permanently trapped in an optical dipole trap. The resulting AC Stark shift of the relevant $\ket{5^2\mathrm{S}_{1/2},F=1} \rightarrow \ket{5^2\mathrm{P}_{3/2},F=1,m_F=\pm1}$ transition (atom A) and of the $\ket{5^2\mathrm{S}_{1/2},F=1} \rightarrow \ket{5^2\mathrm{P}_{3/2},F=1,m_F=0}$ transition (atom B) are 150\,MHz and 115\,MHz, respectively. The atoms are held at the center of a high-finesse optical cavity (finesse $6\times10^4$, mirror distance 0.5\,mm, mode waist radius 30\,\textmu m). Both cavities have asymmetric mirror transmissions of $T_1<6$\,ppm and $T_2\approx 100$\,ppm, leading to a highly directional ($\ge0.9$) single output mode which is matched to a single-mode optical fiber (efficiency typically 0.87). Both systems produce photons on the D$_2$ line of $^{87}$Rb at a wavelength of 780\,nm. In this configuration both atom-cavity systems operate in the intermediate coupling regime of cavity QED (coherent atom-cavity coupling $g\leq 2\pi\times5$\,MHz, cavity field decay rate $\kappa=2\pi\times3$\,MHz, atomic polarization decay rate $\gamma=2\pi\times3$\,MHz).

When a single atom is trapped inside each of the cavities (about 25\,\% of the total measurement time), the experimental protocol runs at a repetition rate of 10\,kHz, including optical pumping (15\textmu s and 4\,\textmu s for node A and B, respectively), photon production for entanglement generation, photonic Bell-state measurement (BSM) and state detection (3\,\textmu s), and optical cooling of atomic motion (70\textmu s at node A, 85\,\textmu s at node B). The necessary laser beams impinge perpendicular to the cavity axis. The presence and position of single atoms is monitored in realtime by collecting atomic fluorescence light on an electron multiplying CCD camera. In combination with a longitudinally shiftable standing-wave dipole trap, the atoms are actively positioned at the center of the cavity mode.

\subsection*{Photon generation}
\begin{figure}[h]
\centering
\includegraphics[width=\columnwidth]{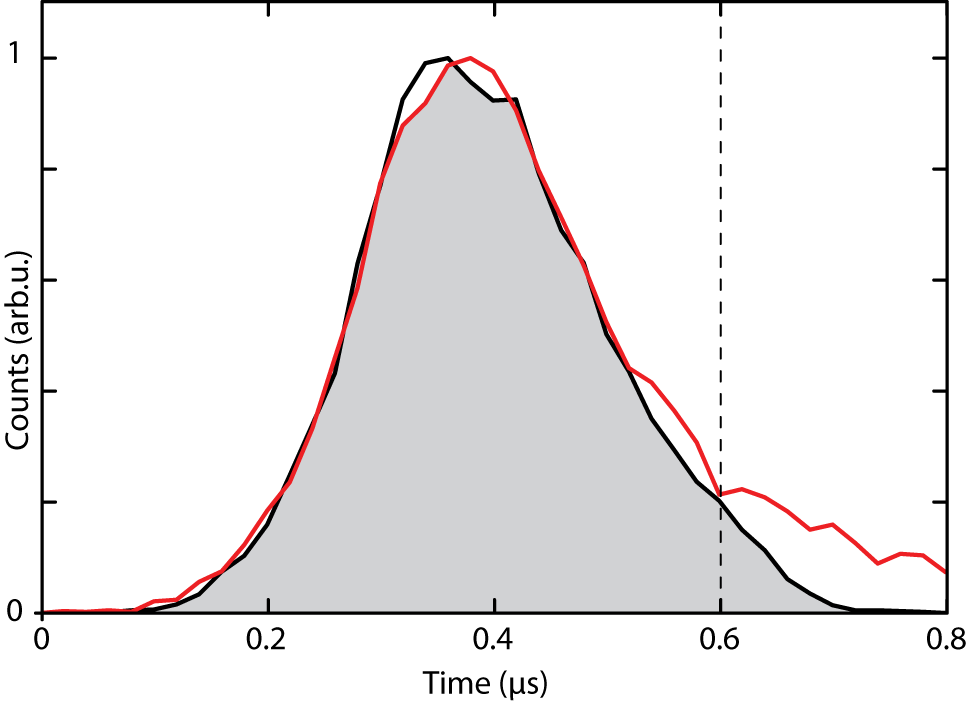}
\caption{
Typical histogram of detector clicks of photon A$^\prime$ (red) generated from atom A and photon C (black) generated from atom B.
\label{fig:methods01}}
\end{figure}

The optical Bell-state measurement requires two-photon interference with high contrast and consequently the possibility to produce two identical photons. Equal frequency and similar temporal profiles are achieved by independent adjustment of the vacuum-stimulated Raman adiabatic passage (vSTIRAP) used to produce the photons. The frequencies of the control lasers in the two setups are the same, as well as the resonance frequencies of the cavities. Control laser and cavity frequency differ by the ground-state hyperfine splitting such that each system is in two-photon resonance. Temporal shapes are adjusted by tuning the Rabi frequencies of the control laser pulses. Figure \ref{fig:methods01} shows histograms of the photon detection times. The photons from node A arriving very late are the result of imperfect optical pumping. The late arrival time of these photons indicates that atomic state preparation and/or atom-to-photon state mapping was not successful. For the teleportation protocol we therefore only consider clicks between $t=0$ and $t=0.6$\,\textmu s (dashed line in Fig.\,\ref{fig:methods01}) to reduce the detrimental influence of these events and wave packet mismatch.

\subsection*{Efficiency of the protocol}
Atom A is prepared in a well-defined state $\ket{\varphi}_\mathrm{A}$ by mapping the polarization of a coherent laser pulse onto the atom. Initially, the atom is prepared in $\ket{F=1, m_F=0}$. When the laser pulse impinges on the cavity, the Rabi frequency of a $\pi$-polarized control laser is ramped down, mapping the polarization onto a superposition of the two atomic states $\ket{F=2, m_F=\pm1}$ \cite{Specht2011}. Using a coherent pulse with on average five photons leads to a measured efficiency of 73\,\%. This value is defined as the probability with which the atom is transferred from $F=1$ to $F=2$ in the storage process. An efficiency of the state preparation smaller than unity does not affect the teleportation fidelity, as in those cases the atom-to-photon state mapping, which is part of the BSM measurement, fails to produce a photon.

After state preparation, single photons are produced with an efficiency $\eta^\mathrm{A}=0.39$ at node A and $\eta^\mathrm{B}=0.25$ at node B. These intracavity photons will leave the cavity through the outcoupling mirror with a probability $T_\mathrm{out}=0.9$. Photons from node A and B are transmitted to the BSM with an efficiency of $T_\mathrm{opt}^\mathrm{A}=0.62$ and $T_\mathrm{opt}^\mathrm{B}=0.24$, respectively. The reasons for the comparatively low transmission from node B are an additional 50:50 beam splitter used to extract photons for state detection at node B and one additional fiber in the optical path. With the quantum efficiency of our single-photon detectors of $\varepsilon\simeq0.55$, the probability to detect an intracavity photon is given by $P_\mathrm{det} = T_\mathrm{out} T_\mathrm{opt} \varepsilon$. We find $P_\mathrm{det}^\mathrm{A} = 0.31$ for node A and $P_\mathrm{det}^\mathrm{B} = 0.12$ for node B. The probability of a detector click per photon-production attempt at node A or B is then $\xi^\mathrm{A}=\eta^\mathrm{A} P_\mathrm{det}^\mathrm{A}=0.12$ and $\xi^\mathrm{B}=\eta^\mathrm{B} P_\mathrm{det}^\mathrm{B}=0.03$, respectively. The main limitations of the total efficiency are transmission losses and imperfect photodetectors ($P_\mathrm{det}$) rather than the photon generation and collection process ($\eta T_\mathrm{out}$). The optical BSM only allows for unambiguous identification of two of the four maximally entangled Bell states. Restricting ourselves to detections of the $\ket{\Psi^-}$ state that directly heralds a successful teleportation event, we find an overall success probability of the teleportation protocol of $\frac{1}{4}\xi^\mathrm{A} \xi^\mathrm{B}=0.09\,\%$. All quoted efficiencies are averaged over all measurements presented in this paper. When an atom is trapped in each of the cavities, the experiment is repeated at a rate of 10\,kHz, which results in one teleportation event every 0.1\,s.

\subsection*{Teleportation fidelity}
The teleportation fidelity depends on the quality of all individual steps of the protocol, in particular on the entanglement generation and the Bell-state measurement. The measured teleportation fidelity will also include imperfections in the preparation of $\ket{\varphi}_\mathrm{A}$ at atom A and in the readout of the final state of atom B. Therefore, the fidelities given in the paper are a lower bound to the fidelity of the teleportation protocol itself.

The process of state preparation and state mapping at node A can be characterized independently by calculating the overlap between the polarization state of photon A$^\prime$, read out from atom A, with the ideal input state: $\Fidelity_\mathrm{A} = \bra{\varphi_\mathrm{A}}\rho_\mathrm{A^\prime}\ket{\varphi_\mathrm{A}}$, with $\rho_\mathrm{A^\prime}$ being the density matrix of the measured state. We find $\FidelityA \approx 95\,\%$ averaged over all input states.
Assuming that non-perfect state preparation and readout at node A results in the state being partially mixed, i.e.
\begin{equation}
\rho_\mathrm{A^\prime} = p_\mathrm{A}\ket{\varphi_\mathrm{A}}\bra{\varphi_\mathrm{A}}+\frac{1}{2}(1-p_\mathrm{A})\mathbbm{1}_2,
\end{equation}
the probability $\pA$ that the read-out state is the ideal state is related to the fidelity by
\begin{equation}
\FidelityA = \frac{1}{2}\left( \pA + 1 \right).
\end{equation}

Similarly, we assume that the entangled state is of the form
\begin{equation}
\rhoEnt = \pEnt\ket{\Psi^-}\bra{\Psi^-}+\frac{1}{4}(1-\pEnt)\mathbbm{1}_4,
\end{equation}
where $\pEnt$ denotes the probability that the read-out atom-photon state is the ideal, maximally entangled Bell state $\ket{\Psi^-}$. In this case, the entanglement fidelity is
\begin{equation}
\FidelityEnt = \bra{\Psi^-}\rhoEnt\ket{\Psi^-} = \frac{1}{4}+\frac{3}{4}\pEnt.
\end{equation}
To determine $\FidelityEnt$, we produce atom-photon entanglement and subsequently map the atomic state onto a second photon \cite{Wilk2007, Weber2009}. Polarization analysis of the resulting two-photon state in three mutually unbiased bases yields an overlap with the maximally entangled $\ket{\Psi^-}$ Bell state of $\FidelityEnt = 89\,\%$.

The quality of two-photon interference also affects the teleportation fidelity and is characterized by the interference contrast $C$, which is the probability that detector clicks signaling a $\ket{\Psi^-}$ detection have been caused by the two photons actually being in a $\ket{\Psi^-}$ state.

Special emphasis has to be put on the fact that the influence of the interference contrast depends on the input state. We discriminate between the four input states $\spindown$, $\spinup$, $\spindowny$, and $\spinupy$ and the two input states $\spindownx$ and $\spinupx$. The latter two are special, because they are mapped onto eigenpolarizations of the detection basis of the Bell-state analyser. In this case, the fidelity is independent of the interference contrast, because classical correlations are already sufficient to signal a successful state transfer---even if the photons did not interfere.
This is clearly pronounced in the teleported state at node B [Fig.\,3(b)] where the corresponding Bloch sphere is elongated along the axis defined by $\spindownx$ and $\spinupx$.

In the following we analyse the teleportation fidelity $\FidelityOrth$ for the four input states $\spindown$, $\spinup$, $\spindowny$, and $\spinupy$. From the definition of the state fidelity one finds
\begin{equation}
\FidelityOrth=1-\pw,
\label{eq:StateFidelity}
\end{equation}
where $\pw$ is the probability for a wrong result during state detection. The probability for the protocol to succeed upon detection of a $\ket{\Psi^-}$ correlation is $C\pEnt\pA$. We assume that whenever the teleportation protocol fails in spite of detection of a $\ket{\Psi^-}$ correlation (because of no interference or no entanglement or incorrect state preparation), the result of the state detection at node B is independent of the state we intended to prepare at node A, resulting in a fidelity of $1/2$.
This yields
\begin{equation}
\pw = \frac{1}{2}\left( 1 - C\pEnt \pA \right).
\end{equation}
Using Eq.\,(\ref{eq:StateFidelity}) we find:
\begin{equation}
\FidelityOrth = \frac{1}{2} + \frac{1}{2} C\pEnt\pA.
\label{eq:FidelityTheory}
\end{equation}
Using the expressions for $\pEnt$ and $\pA$, this can be written as
\begin{equation}
\FidelityOrth = \frac{1}{2} + \frac{4}{3} C \left(\FidelityEnt-\frac{1}{4}\right)\left(\FidelityA-\frac{1}{2}\right).
\label{eq:FidelityTheory1}
\end{equation}
An analogous analysis for eigenpolarizations of the Bell-state analyser leads to
\begin{equation}
\FidelityBell = \frac{1}{2} + \frac{4}{3} \left(\FidelityEnt-\frac{1}{4}\right)\left(\FidelityA-\frac{1}{2}\right).
\label{eq:FidelityTheory2}
\end{equation}
Equations\,(\ref{eq:FidelityTheory1}) and (\ref{eq:FidelityTheory2}) are plotted in Fig.\,3(a) as a dashed blue and a dashed black line, respectively, using $\FidelityEnt = 89\,\%$ and $\FidelityA = 95\,\%$.

Averaging over Eq.\,(\ref{eq:FidelityTheory1}) and (\ref{eq:FidelityTheory2}) for the six above mentioned input states readily results in
\begin{equation}
\Fidelity = \frac{1}{2} + \frac{8}{9} \left( C + \frac{1}{2} \right) \left( \FidelityEnt - \frac{1}{4} \right) \left( \FidelityA - \frac{1}{2} \right).
\label{eq:FullFidelityTheory}
\end{equation}
Using independently measured values for $\FidelityEnt$ and $\FidelityA$ and therefore without any free parameter, this simple model shows good agreement with the experimental data (Fig.\,3).

\end{document}